\documentclass{article}

\usepackage{arxiv}

\usepackage{abbrevs}
\usepackage{wrapfig}
\usepackage[utf8]{inputenc} 
\usepackage[T1]{fontenc}    
\usepackage{hyperref}       
\usepackage{url}            
\usepackage{booktabs}       
\usepackage{amsfonts}       
\usepackage{nicefrac}       
\usepackage{microtype}      
\newcommand{\multilinecomment}[1]{}
\usepackage{graphicx,psfrag,amsmath,amsfonts,verbatim,mathtools,bbm,tikz,bm}

\newtheorem{theorem}{Theorem}

\newtheorem{definition}[theorem]{Definition}
\newtheorem{criterion}[theorem]{Criterion}

\newtheorem{example}[theorem]{Example}

\DeclareMathOperator*{\argmax}{arg\,max}

\newcount\Comments  
\definecolor{darkgreen}{rgb}{0,0.5,0}
\newcommand{\kibitz}[2]{\ifnum\Comments=1\textcolor{#1}{#2}\fi}
\newcommand{\adigi}[1]{\kibitz{blue}{[AD: #1]}}
\newcommand{\ezell}[1]{\kibitz{darkgreen}{[EZ: #1]}}

\newcommand{\calA}{\mathcal{A}}
\newcommand{\calB}{\mathcal{B}}
\newcommand{\calP}{\mathcal{P}}
\newcommand{\calS}{\mathcal{S}}
\newcommand{\calC}{\mathcal{C}}
\newcommand{\calO}{\mathcal{O}}
\newcommand{\calM}{\mathcal{M}}
\newcommand{\N}{\mathbb{N}}
\newcommand{\EE}{\mathbb{E}}
\newcommand{\R}{\mathbb{R}}

\newcommand{\bba}{\boldsymbol{a}}
\newcommand{\bbpi}{\boldsymbol{\pi}}
\newcommand{\eps}{\varepsilon}

\newname\mal{Multi-agent learning (MAL)}[MAL]
\newname\mdp{Markov decision process (MDP)}[MDP]
\newname\ebs{egalitarian bargaining solution (EBS)}[EBS]

\title{Survey of Self-Play in Reinforcement Learning}

\author{%
  Anthony DiGiovanni$^1$ and Ethan Zell$^2$ \\
  Departments of Statistics$^1$ \& Mathematics$^2$\\
  University of Michigan\\
  \texttt{\{adigi,ezell\}@umich.edu} \\
}

\begin{document}

\maketitle

\begin{abstract}
  In reinforcement learning (RL), the term self-play describes a kind of multi-agent learning (MAL) that deploys an algorithm against copies of itself to test compatibility in various stochastic environments. As is typical in MAL, the literature draws heavily from well-established concepts in classical game theory and so this survey quickly reviews some fundamental concepts. In what follows, we present a brief survey of self-play literature, its major themes, criteria, and techniques, and then conclude with an assessment of current shortfalls of the literature as well as suggestions for future directions. 
\end{abstract}

\multilinecomment{
\begin{itemize}
    \item outline:
    \begin{itemize}
        \item \ezell{1) Nash-based}
        \item \ezell{2) coop, Pareto + EBS (unique, fair), learning eq? (examples to highlight differences in criteria), punishment for deviants (FCL and P+S)}
        \item \adigi{open problems: getting both self play and adaptability (esp w/ unknown other player rewards)}
    \end{itemize}
\end{itemize}
}

\section{Introduction and Background}\label{sec:intro}


Multi-agent learning (MAL) research aims to study interacting systems of agents in a shared environment, the incentives that govern the agents' behavior, and the optimal way to behave in such environments. As we will see, the definitions of ``optimality,'' ``environment,'' and even ``incentive'' are crucial to developing useful problem statements and powerful solutions. 

Several agendas with distinct objectives and evaluation criteria have been developed for MAL \cite{question}. From the perspective of online and reinforcement learning (RL), it is natural to formulate MAL as design of algorithms that can achieve a user's goals optimally (or minimize \textit{regret}), conditional on an environment induced by the actions taken by other algorithms. Model classes are constructed for this induced environment, on a spectrum from fully stochastic and equivalent to a Markov decision process (MDP) \cite{ICML10-chakraborty}, to fully adversarial and unpredictable \cite{rmax}. Alternatively, game theoretic approaches typically ask how to converge to an equilibrium, assuming play by other agents that is ``rational'' in some sense. For the design of an algorithm that multiple users will want to adopt, a guiding principle is \textit{self-play compatibility} (or simply ``compatibility''): achieving some high performance benchmark when copies of the algorithm play each other \cite{PS04}. Maintaining such performance while also adapting to non-self agents, in line with the RL agenda, presents a challenging research program.

To appraise the state of compatibility research, particularly theoretical results, we present a survey of literature on this problem and conjecture some implications of these works. Our survey has revealed a number of limitations, open problems, and themes.

First, even within the narrow MAL subproblem of compatibility, different subsets of the literature pursue disparate definitions of ``good'' performance (detailed in Section \ref{sec:criteria}). These range from the relatively simple property of convergence to any Nash equilibrium of the stage game \cite{ ICML10-chakraborty, CS06}, to the more demanding criteria of Pareto efficiency \cite{PS04} in an equilibrium of the stage or repeated game, and fairness in the senses studied in cooperative bargaining theory \cite{TD20}. We argue that research targeted at these latter criteria is practically essential, given that multiple users of a MAL algorithm will want it to achieve the highest rewards possible within the constraints of cooperation. 

Second, finite-time guarantees (particularly regret bounds) of self-play performance are almost nonexistent, and the exceptions \cite{sharpSelfPlay, TD20} are not designed or proven to maintain high rewards against simple classes of non-self players. While several works prove polynomial-time convergence, losses incurred during the time it takes for self-copies to coordinate on an equilibrium may be significant.


Third, most existing algorithms with both provable self-play compatibility, and success against other classes of players, use an approach that can be summarized as: ``propose, then adapt.'' 
The proposal is the solution that all self copies eventually coordinate on, such as a Nash equilibrium. At least when the preferences of the other agents are known, satisfying any single property desired for MAL algorithms is not too demanding. The challenge lies in simultaneously maintaining multiple guarantees.
Thus, many compatible algorithms identify a target class of non-self players to adapt to, in the event that an initial testing period gives evidence that the other player is not a copy. 


\section{Research Aims}

This survey explores and conjectures answers for the following questions:
\begin{itemize}
    \item Which formal models of self-play are most theoretically tractable, and what proof techniques and tools are most commonly used? (For example, how do proofs of convergence in self-play \cite{ICML10-chakraborty, CS06} overcome the difficulty of adaptive nonstationarity, which depends on the actions of each copy of the algorithm?)
    \item Because ``optimal'' self-play does not appear well-defined, what are the standard benchmarks for evaluating an RL algorithm's performance in self-play, especially in regret definitions? Which computational tradeoffs might exist for higher standards, e.g.\ achieving Pareto efficient solutions \cite{satisficing,TD20} versus any Nash equilibrium \cite{CS06}?
    \item How do certain algorithms \cite{ICML10-chakraborty, boundedMem, TD20} maintain self-play compatibility while also guaranteeing safety against adversarial players, and optimality against stochastic ones?
\end{itemize}

Our purpose is a focused treatment of self-play theory, rather than a broad list of summaries of the many papers in this field. Omission of a given work is therefore not an implicit judgment of the lack of importance of that work.

\section{Key Results}

\subsection{Background:\ Markov Games}


Markov games appear as an extension of usual game theory since they incorporate some element of randomness or hidden values to a game theoretic scenario. Since Markov games involve two or more deciding agents, they also extend the notion of an MDP. For our purposes, a Markov game will adhere to several simplifying assumptions. As usual in reinforcement learning, the game's players interact with the environment by choosing actions from a (finite) set $\calA_i$ and receiving a (possibly random) reward. Games are also repeated. Moreover, Markov games are often assumed to have infinite or unknown horizon with ``enough'' time for players to learn the game. 

For simplicity, we discuss games that progress in discrete steps $t\in \N$, though it is possible to adapt many of these techniques for a continuous time analogue. For player $i$ at time $t$, we will denote $R^{i}_t$ to be the reward. Some papers view this reward as deterministic \cite{satisficing}, while others view it as the distribution of a random variable. In the case the reward is random, players only receive a sample from the distribution and therefore must make estimates as to what the distribution actually is. Often, $R^i_t$ will depend not only on the $i^{th}$ player's action, but also on the other players' actions as well. To emphasize this, we may write $R^i_t(\boldsymbol{a})$ where $\bba \in \calA := \prod_i \calA_i$. We will also further restrict ourselves to the case where there are finitely many players, $\{1,\dots,n\}$, and finitely many states, $\{1,\dots,d\}$. For brevity, we will denote these sets by $[n]$ and $[d]$, respectively. 

\begin{definition}
A \textbf{strategy} or \textbf{policy} $\pi_i$ for player $i$ is a map that takes as inputs the game stage and player's possible positions to the set of probability distributions over actions:
\[
\pi_i : \N\times [d] \to \calP(\calA_i).
\] If the strategy $\pi_i$ is independent of time, we refer to it as a \textbf{stationary strategy}. We will denote the set of all strategies for player $i$ by $\calS_i$. One could also define a strategy that depends on the positions of all players.

\end{definition}

 In words, $\pi_i$ tells the $i^{th}$ player how to behave given its own state and the stage of the game. Player $i$ will then choose action $a \in \calA_i$ with probability $(\pi_i)_a$. Note that this definition includes deterministic strategies since one could specify the probability of a particular action to be $1$.
 
 \begin{definition}
 We define a \textbf{strategy profile} $\bbpi$ as the map which contains the information on the strategies of all players:
\[
\bbpi := \pi_1\times\cdots\times\pi_n : \N\times [d]^{[n]} \to \prod_{i\in [n]}\calP(\calA_i).
\]
 \end{definition}

In several of the papers we discuss, it is allowed that players have access to different actions \cite{sharpSelfPlay}. Still, an important special case is called a \textit{repeated symmetric game} (RSG). 

\begin{definition}
A repeated Markov game is called \textbf{symmetric} if, for any strategy profile $\bbpi$ and any permutation of the players $\sigma \in S_n$,
\[
R^{\sigma(i)}(\pi_{-i},\pi_i) = R^i(\pi_{\sigma(-i)},\pi_{\sigma(i)})
\] for all $i\in [n]$. By $\pi_{-i}$ we mean the profile $\bbpi$, but with the $i^{th}$ entry omitted. Similarly, by  $\sigma(-i)$ we mean the permutation $\sigma$ applied to all players except the $i^{th}$.
\end{definition}

In both symmetric and general games, the concept of \textbf{Nash equilibrium (NE)} arises naturally. More generally, 

\begin{definition}
Let $\eps\geq 0$. A strategy profile $\bbpi^*$ is called a $\eps$-NE if for all strategies $\pi_i \in \calS_i$:
\[
\EE [ R^i_t(\pi_i, \pi_{-i}^*)] \leq \EE [R^i_t(\bbpi^*) ]+\eps,
\] for all players $i\in[n]$. When $\eps =0$, this is the usual notion of NE.
\end{definition}

In classical game theory, NE do an excellent job of assessing when players have no incentive to deviate from established behavior. Unfortunately, NE may be sub-optimal. For instance, consider the Stag Hunt game (Table \ref{tab:sh}). When both players cooperate and hunt the stag, both receive higher reward than all other actions in the game. However, if one player defects and hunts the rabbit instead, then neither succeeds at hunting the stag alone. Here, we can check there are two NE: $(\text{Stag},\text{Stag})$ and $(\text{Rabbit},\text{Rabbit})$. Yet, $(\text{Rabbit},\text{Rabbit})$ results in only a fraction of the maximum reward. 

Hence, we introduce alternative notions to test for optimality.

\subsection{Selected Evaluation Criteria in Multi-Agent Learning}\label{sec:criteria}


The literature on self-play features various solution concepts. We present both the self-play criteria and complementary criteria for other-play. Let $\mathfrak{A}$ be the algorithm under consideration, and player 1 be the agent using $\mathfrak{A}$. A \textit{stage game} is the game corresponding to one state of the full Markov game, given by $R^i(\textbf{a})$ for each $i$ (for example, the bimatrix of Stag Hunt is a stage game). The most basic property of interest \cite{ICML10-chakraborty, CS06} is Stage Game Convergence:

\begin{criterion}[Stage (Markov) Game Convergence] With high probability, the policies of players who all use $\mathfrak{A}$ converge to an $\eps$-Nash equilibrium of the stage (Markov) game.
\end{criterion}

Both versions of this criterion are properties of the algorithm's output \textit{policies}, not of the empirical distribution of actions. Despite its simplicity, Stage Game Convergence is stronger than convergence of the empirical distribution to a stage game Nash equilibrium, which is not even guaranteed in all general-sum games by the classic algorithm of fictitious play \cite{shapley}.

Powers and Shoham \cite{PS04} refined this criterion to focus on \textit{rewards} of the algorithm in self- and other-play. To understand these properties, we need the following notions. A strategy profile $\boldsymbol{\pi}$ is $\eps$-\textit{Pareto dominated} if there exists some $\boldsymbol{\pi'}$ such that $\mathbb{E}_{\boldsymbol{\pi'}}(R^i) \geq \mathbb{E}_{\boldsymbol{\pi}}(R^i) + \eps$ for all $i$, and $\boldsymbol{\pi}$ is $\eps$-\textit{Pareto efficient} if it is not $\eps$-Pareto dominated.

\begin{criterion}[PS-Properties]\label{psProps} Let $k$ be the number of outcomes of the game, and let $b := \max_{i=1,\dots,n} \{\max_{\textbf{a}} R^{i}(\textbf{a}) - \min_{\textbf{a}} R^{i}(\textbf{a})\}$. For any $\eps, \delta > 0$, there exists a polynomial $p$ and a quantity $T_0 := p\left(\frac{1}{\eps}, \frac{1}{\delta}, k, b\right)$  such that, for any game length $T > T_0$, the algorithm achieves an average reward of at least the following benchmarks, with probability at least $1-\delta$:
\begin{itemize}
    \item \textbf{Targeted Optimality:} Let $\mathcal{C}$ be a target class of opponents, $\mathfrak{B} \in \mathcal{C}$ be the actual opponent's algorithm, and $V_{BR} := \max_{\mathfrak{A'}} \lim_{t \to \infty} \frac{1}{t} \sum_{j=1}^t \mathbb{E}_{\mathfrak{A'},\mathfrak{B}}(R^{1}_j)$ be the value of the best response to the opponent (in the space of algorithms, not actions of the stage game). Against $\mathfrak{B}$, the average reward of $\mathfrak{A}$ is at least $V_{BR} - \eps$.
    \item \textbf{Stage (Markov) Game Pareto:} Let $V_{P}$ be the minimum expected value to the player from the set of $\eps$-Pareto efficient Nash equilibria of the stage (Markov) game. Against $\mathfrak{A}$, the average reward of $\mathfrak{A}$ is at least $V_{P} - \eps$.
    \item \textbf{Safety:} Let $V_S^1 := \max_{\pi_{1}} \min_{\pi_{-1}} \mathbb{E}_{\bbpi} (R^{1})$ be the security value. Against any opponent, the average reward of $\mathfrak{A}$ is at least $V_S^1 - \eps$.
\end{itemize}
\end{criterion}

Stage Game Pareto provides an intuitive ordering of NEs; if all players are at least $\eps$ better off in one NE than in another, the former is preferable. In some games (see Section \ref{sec:stageconv}), such as the Prisoner's Dilemma (Table \ref{tab:pd}), even Stage Game Pareto is too weak. This motivates Markov Game Convergence, which encompasses the case of repeated games if there is only one stage game.

The concept of regret in self-play evaluation does not translate easily from online learning. The strongest notion of regret compares to the optimal rewards $\mathfrak{A}$ could have received over all possible sequences of actions, conditional on the sequence of actions the other players would have used in response to our learner's actions \cite{policyreg}. This counterfactual makes analysis in self-play very challenging, since it is unclear how to characterize the optimal response to one's own algorithm, which in general produces highly nonstationary behavior that is a function of the entire game history. In adversarial online learning problems modeled as zero-sum games against nature, previous works responded to this difficulty by considering the more tractable benchmark of \textit{external} regret, that is, comparing to the best fixed action \cite{external}. Considering that in self-play one can design an algorithm to be cooperative, rather than adversarial or completely unpredictable, such a relaxation is too weak for this setting. 

A recent work \cite{TD20} thus considers a compromise between these extremes:\ compare the algorithm's rewards to a fixed value, which is Pareto efficient in the \textit{repeated} game and satisfies some property of fairness. Their chosen target is the egalitarian bargaining solution (EBS), defined as follows.

\begin{definition} Let $V_S^i$ be player $i$'s security value, as defined in Criterion \ref{psProps}, and in a two-player game let $\mathcal{G} := \{(R^{1}(i,j),R^{2}(i,j)) \ | \ i \in \mathcal{A}_1, j \in \mathcal{A}_2\}$. Let the set of feasible and individually rational rewards be $\mathcal{U} := \text{Conv}(\mathcal{G}) \cap \{(u_1, u_2) \ | \ u_1 \geq V_S^1, u_2 \geq V_S^2\}$. For pairs $x, y \in \mathcal{U}$, define the lexicographic ordering $\geq_\ell$ by: $x \geq_\ell y$ iff (1) $\min\{x_1, x_2\} > \min\{y_1, y_2\}$ or (2) $\min\{x_1, x_2\} = \min\{y_1, y_2\}$ and $\max\{x_1, x_2\} \geq \max\{y_1, y_2\}$. Then, with $V_S := (V_S^1, V_S^2)$ and $V(\boldsymbol{\pi})^i := \lim_{t \to \infty} \frac{1}{t} \sum_{j=1}^t \mathbb{E}_{\boldsymbol{\pi}}(R_j^i)$, a joint policy $\boldsymbol{\pi}_{Eg}$ is an \textbf{EBS} if $V(\boldsymbol{\pi}_{Eg}) - V_S \geq_\ell V(\boldsymbol{\pi}) - V_S$ for all $\boldsymbol{\pi}$. The EBS values are the pair $V_{Eg} = V(\boldsymbol{\pi}_{Eg})$.
\end{definition}

That is, this value maximizes the minimum gain of a player over their security value, and ties are broken by the maximum gain. The resulting criterion is then:

\begin{criterion}[Individual Rational Regret] Over $T$ rounds of self-play, with high probability the individual rational regret is sublinear in $T$, that is:
\begin{align*}
    \max_{i=1,2} \sum_{t=1}^T (V_{\text{Eg}, i} - R^{i}_t) &= o(T)
\end{align*}
\end{criterion}

An analogous criterion of safety regret is defined for $V_S$ instead of $V_{\text{Eg}}$, with respect only to player 1 rather than the maximum of both players. Finally, one of the strongest criteria is achievement of Nash equilibrium in the space of \textit{learning algorithms} \cite{fcl}:

\begin{criterion}[Learning Equilibrium] For each game $g$ in some set $\mathcal{G}$, there exists a time $T_g$ such that, with respect to expected limit-average rewards after $T_g$, use of $\mathfrak{A}$ by all players is a Nash equilibrium of the game whose action space is the set of learning algorithms.
\end{criterion}

This objective is appealing because, in contexts where the game dynamics are unknown and thus need to be learned online, it is not ideal to assess the stability of a self-play outcome by considering unilateral deviations of \textit{policies}. The relevant question is instead whether any user of the learning algorithm itself has an incentive to switch to another algorithm.

In the following subsections, we will discuss key papers aimed at each of these criteria.

\subsection{Stage Game Convergence, Targeted Optimality, and Safety}\label{sec:stageconv}

Once a common NE is found, algorithms like AWESOME \cite{CS06} and Convergence with Model Learning and Safety (CMLeS) \cite{ICML10-chakraborty} look to achieve stage game convergence. In AWESOME's case, this is achieved by updating two hypotheses on a carefully spaced \textit{schedule}. First, AWESOME would like to play optimally against stationary opponents and maintains the null hypothesis that opponents are playing stationary strategies, that is, fixed distributions over actions not conditioned on the past. Second, AWESOME maintains a null hypothesis that its opponents are playing a NE. After each stage of the game, AWESOME uses maximum likelihood estimates to possibly reject either of these hypotheses. 

AWESOME's cycle of play proceeds as follows. Until AWESOME rejects that its opponents are stationary, it will play its best possible strategy to exploit them. If AWESOME rejects this hypothesis, it will retreat to playing a pre-computed NE. Once AWESOME rejects that the opponents are playing the same NE, it will randomly pick a strategy the following round (this helps with exploration). At this point, AWESOME resets and begins the process over. It turns out that a clever choice of \textit{schedule}---that is, a combination of epochs of the algorithm, and thresholds for comparison of estimated strategies to test hypotheses---allow provable convergence to a best response against stationary opponents and provable convergence to NE in self-play, both with probability $1$.

CMLeS builds on the approach of AWESOME, attempting to nearly satisfy the PS-Properties. The self-play objective is still Stage Game Convergence. However, this is balanced with a guarantee of safety, and the class for Targeted Optimality is extended to opponents that condition actions on past joint actions (``bounded memory''). The former follows by retreating to the strategy $\argmax_{\pi_1} \min_{\pi_{-1}} \mathbb{E}_{\bbpi} (R^{1})$ whenever rewards drop below $V_S^1 - \eps$. To achieve the latter, if CMLeS discovers that at least one other player is not a self-copy, it switches to a model-based RL algorithm, R-MAX \cite{rmax}, taking the past actions as the state of an MDP. Since, unlike AWESOME, each CMLeS player does not necessarily start playing the same stage game NE, the copies communicate their identities by playing a sequence of actions that a bounded memory player would not always match.

Optimizing against bounded memory opponents lets a CMLeS player account for adaptations by other players to the past. To see how this could be useful, consider the Prisoner's Dilemma (Table \ref{tab:pd} in the Appendix). If P2 always plays the same action as P1's most recent action, P1 can maximize average rewards by playing this  policy as well. Incorrectly modeling P2 as stationary would prescribe always playing Defect, the (suboptimal) stage game NE. This example highlights, however, the weakness of CMLeS's Stage Game Convergence goal; the stage game NE gives reward 1, rather than the cooperative reward of 3 that could be achieved with the policy constructed above (a repeated game NE). The authors' strategy for proving convergence\textemdash repeated sampling from the set of NEs\textemdash would not work for Markov game NEs, because infinitely many such NEs exist (Proposition 144.3 of Osborne and Rubinstein \cite{folkthm}). We also note that even under polynomial-time convergence to NE, the regret with respect to that NE of CMLeS and AWESOME is not bounded by a sublinear rate.

\subsection{Pareto Efficiency}\label{sec:pareto}


Manipulator \cite{boundedMem} uses a similar approach to CMLeS in two-player games: starting with a strategy conducive to its self-play goal, switching to a best response to a bounded memory player if the first strategy fails, and applying a safe strategy when rewards get too low. However, instead of a stage game NE, the initial strategy is as follows:\ play the user's half of a joint strategy that maximizes the user's rewards, subject to the constraint of giving the other player at least their security value. The constraint gives even non-self players an incentive to play the proposed joint strategy, and by design the joint strategy is Pareto efficient in the repeated game. Stage Game Pareto was achieved by a simpler algorithm from the same authors \cite{PS04}. Theoretically, the source of Manipulator's guarantees also strongly resembles CMLeS, tuning epoch lengths and the probability of switching to the best response strategy to balance the three criteria. 

The S-Algorithm provably reaches Markov Game Pareto in most repeated games \cite{satisficing}. It does not guarantee Targeted Optimality or Safety, unlike Manipulator. However, its key advantage is that it does not need knowledge of the other players' rewards. Multiple agents who may want to keep their reward functions private, to avoid exploitation in other-play for example, can thus achieve Pareto efficiency. A limitation of this algorithm is that, as explored in Section \ref{sec:learneq}, a Pareto efficient outcome is not necessarily fair.

The S-Algorithm uses the principle of adaptive satisficing:\ each agent starts with an aspiration level, 
updates the aspiration towards the reward received at each time step, and repeats the most recent action if its reward was at least the most recent aspiration. 
The proof of convergence to $\eps$-Pareto efficiency relies on the following observations. If at least one player's aspiration falls below its Pareto value $V_P$, a non-Pareto efficient action may exceed this aspiration and be reinforced by the satisficing rule. However, suppose the players all initialize their aspirations higher than the Pareto values. Then a lower bound can be shown for the number of time steps in which the only joint actions that meet the players' aspirations are $\eps$-Pareto efficient, and at least one such joint action exists. Therefore, with sufficiently slow learning rates for aspiration updates, the self-copies can repeat their independent draws of exploratory actions many times, and consequently control the probability that they never find this $\eps$-Pareto efficient solution\textemdash which, once found, is maintained for the rest of the game.

This proof strategy resembles that of AWESOME and CMLeS, in that the self copies discover a compatible joint strategy simply by repeating uniform random sampling, and the agents' learning rates must be tuned to avoid deviations from the solution. While the S-Algorithm does not need to know the other players' rewards or collect statistics on their strategies like the former, substituting inequality 5 into Lemma 3 reveals that its time complexity for convergence is still exponential in $n$. 

We highlight that lack of full knowledge of other players' rewards, or \textit{incomplete information}, fundamentally increases the difficulty of achieving compatibility. This restriction makes it impossible to compute a Pareto efficient solution at the start, or to straightforwardly punish deviations from this solution (the player does not know which actions constitute ``punishment''). 
This explains the relative dearth of works showing both compatibility and other-play optimality in games of incomplete information. Some exceptions that currently lack substantial theoretical analyses include S++ \cite{C14}, which modifies the S-Algorithm to select among sub-algorithms rather than primitive actions, and M-Qubed \cite{CG10}, which combines Q-learning with optimistic value estimates and a safety backup policy.



\subsection{Markov Game Convergence}\label{sec:markovconv}

Since algorithms like AWESOME assume the calculation of a common Nash Equilibrium between its self-copies, this begs the question: ``how can one efficiently compute a NE?'' In an early work of this kind, \cite{bayes} shows that Bayesian updating in repeated games results in players learning a repeated game NE, even under incomplete information. In the early model, players are assumed to have perfect-monitoring of the outcomes in an infinitely repeated game with discounting. Bayesian updating works by retaining estimates for the payoffs of the game and the strategies of the other players. By doing so, rational agents can learn to play optimally against stationary opponents or otherwise converge to an $\eps$-NE.

Crucially, \cite{bayes} was among the first results to expand beyond myopic theories of simultaneous learning; that is, they do not assume that their opponents actions are fixed and allow for the opponents also to change and learn. As it is an early iteration, \cite{bayes} makes a strong absolute continuity assumption related the empirical measure and the ``true'' measure of opponents' actions. In doing so, they gain access to powerful tools from classical analysis like the Radon-Nikodym theorem, but raise the question of whether this assumption is too strong.

Many other algorithms find ways of computing NE and the scope of this question is beyond a brief survey. However, we will mention \cite{sharpSelfPlay} since it is the current best complexity for finding an $\eps$-NE in a general game. In \cite{sharpSelfPlay}, the authors introduce a new model-based algorithm Multiplayer Optimistic Nash Value Iteration (Multi-Nash-VI) that achieves a complexity of $\tilde \calO(d^2H^4(\prod_{i=1}^n |\mathcal{A}_i|)/\eps^2)$ for finding an $\eps$-NE. Here, $H$ is the length of an episode, that is, a repetition of the Markov game. Of course, the product of the actions in the Multi-Nash-VI bound is an exponentially bad issue as the number of players gets large. An information theoretic lower bound for \textit{zero-sum} two-player games is $\Omega (dH^3 (|\calA_1| + |\calA_2|)/\eps^2)$, though this may not be tight for general-sum games.

Briefly, Multi-Nash-VI works in two stages that in a broad sense are emblematic of many algorithms which compute $\eps$-NE. The algorithm begins with some empirical estimates of the game's parameters and computes a greedy policy with respect to them, using a kind of optimistic value iteration. The second stage is routine: Multi-Nash-VI plays the policy it obtained, collects data from that round of play, and updates its estimates. Philosophically, this is similar to the Bayesian updating of \cite{bayes}---the clever use of bonus parameters and optimistic estimates seems to make all the difference.

\subsection{Individual Rational Regret and Learning Equilibrium}\label{sec:learneq}

First introduced in \cite{originalEBS}, the EBS is one possible improvement on the lackluster NE target for repeated games. In \cite{TD20}, the EBS is exploited for two useful properties: higher expected yield than NE and uniqueness. In contrast, recall that NE (even Pareto efficient ones) need not be unique and for the AWESOME algorithm, one burdensome assumption was that copies of the algorithm compute the same NE. In practice, this may be impossible to guarantee. Crucially, \cite{TD20} makes use of the repeated game's structure to elucidate a strategy for approaching the EBS, even though the action space is finite (and thus not convex like in the original formulation in \cite{originalEBS}).

The EBS is slightly more involved than NE and is best understood through an example. 

\begin{example}
\end{example}

Suppose two friends $A$ and $B$ are playing an asymmetric game with the following possible outcomes:
\[
\{(0,0),(1,0),(0,1),(0,5)\}.
\] In a bargaining problem, these two friends must agree on one outcome or else they will default to the worst possible reward: $(0,0)$. On one hand, $A$ does not have the same bargaining ability as $B$, because $B$ has a much higher possible payoff. However, $B$ also has much more to lose and so the threat of $A$ defecting is challenging. To solve this, the players should not settle on any single action, but rather a probability distribution over actions that both players then commit to. Abstractly, we consider the convex hull $\calC$ of these points in $\R^2$ and define the EBS value as $V_{Eg}(\calC)\in\calC$ where $V_{Eg}$ is defined previously (Section \ref{sec:criteria}). For the example, see Figure \ref{EBS} in the Appendix: $\calC$ is the region in Q1 enclosed by the blue line and the EBS is the labelled point. The egalitarian solution satisfies a strong monotonicity condition in the sense that all players should benefit from any expansion of opportunities, regardless of whether the expansion is biased in favor of one particular player. Precisely, the two main properties of the alternative solution concepts are:

\begin{theorem}
    Let $\calM = (\prod_{i\in [n]}R^i,\prod_{i\in [n]}\calA_i,[d])$ be the information for a repeated game, and let $V_{Eg}$ be the EBS value of the game. Suppose that for some player $i\in [n]$ $\calA_i \subseteq \calB$ and define a new repeated game by the information $(\prod_{i\in [n]}R^i,\prod_{j\neq i}\calA_j \times \calB,[d])$. Let $V_{Eg}'$ be the EBS value of the new game. Then, the EBS solution has the ``monotone'' property $V_{Eg} \leq V_{Eg}'$. 
    
    Alternatively, suppose $\bbpi$ is a Nash bargaining solution \cite{NBS} for $\calM$ and that $\bbpi'$ is a Nash bargaining solution for $((aR^i + b) \times \prod_{j\neq i}R^j,\prod_{i\in [n]}\calA_i , [d])$, where $a,b \in \mathbb{R}$. Then, $\bbpi = \bbpi'$.
\end{theorem}

It is important to note that an EBS solution is not invariant under affine transformations and the Nash bargaining solution is not monotone. Moreover, \cite{originalEBS} proved that a solution satisfies weak Pareto efficiency, symmetry, and strong monotonicity if and only if it is the EBS solution. 

To adapt this bargaining solution notion to repeated games, \cite{TD20} introduces the UCRG algorithm which begins with optimistic exploration. As it collects estimates from its environment, the algorithm produces a collection of games $\calM_t$ ``close'' to the one it is playing in the sense 
\[
\calM_t := \{R \mid |\EE R^i(\bba) - \bar R^i_t(\bba)|\leq C_t(\bba) \quad \forall i,\bba\}
\] where $R$ is the reward structure of the game, $C_t$ is some parameter that vanishes as $t\to\infty$, and $\bar R^i$ is the empirical estimate of the rewards. Heuristically, these are confidence intervals for the game. The UCRG algorithm then behaves optimistically in the sense that it assumes until an update that it is playing the game from $\calM_t$ with the best rewards.
\multilinecomment{
Accompanying this setup, \cite{TD20} introduces the following notion:
\begin{definition}
Let $V_S^i$ be the safety value defined in Criterion \ref{psProps} with respect to the $i^{th}$ player. The \textbf{safety regret} for an algorithm $\Lambda$ played by $i$ for $T$ rounds against opponents using strategies $\pi_{-i}\in\calS_{-i}$ is:
\[
\text{Regret}_T(\Lambda, \pi_{-i}) := \sum_{t=1}^T (V_S^i - R^i_t).
\]
\end{definition}

The upshot is: the UCRG algorithm from \cite{TD20} achieves sub-linear (approximately $\calO(T^{2/3})$) safety regret by optimistically exploring and then playing a policy for the EBS value. \adigi{If we're just reporting the $\calO(T^{2/3})$ rate here from UCRG, as opposed to $\calO(T^{1/2})$ from the optimistic maximin, might be superfluous to define it formally. The EBS by definition is a stronger criterion than safety in self-play, so as the authors note the safety reg guarantee is an immediate corollary - arguably not interesting in self-play because why would self-copies be adversarial against each other? That said I do think safety regret per se (against arbitrary players) is important, just trying to be mindful of space here} \ezell{We can cut}
}



Another recent work has connected this EBS perspective on MAL to Learning Equilibrium. 
Jacq et al.\ \cite{fcl} consider an episodic Markov game, that is, repetition of a Markov game (with more than one state) that terminates after finitely many steps. We will call each repetition an \textit{episode}. This game is also assumed to be symmetric, from which it follows that players can optimize the minimum of all players' values (ignoring the security value due to symmetry) by sequentially permuting, at the start of every episode, player indices of a strategy profile that optimizes their sum.

The principle of their algorithm, Foolproof Cooperative Learning (FCL), is similar to FolkEgal \cite{folkegal}, which constructs a repeated game NE by playing an approximation to the EBS and punishing a player who deviates from the EBS. Unlike FolkEgal, FCL works for Markov games where the dynamics are initially unknown by all players. As in the bandit case of Tossou et al.\ \cite{TD20}, this poses a problem of balancing exploration with reward maximization, which motivates the combination of Q-learning with an exploration schedule in FCL. However, to achieve Learning Equilibrium, FCL goes beyond trying to achieve the EBS in self-play; it must disincentivize players who attempt to use algorithms that do not pursue the EBS, to avoid being cheated.

FCL's solution to this dilemma is to have each user of the algorithm maintain action-value estimates with respect to several surrogate reward functions: (1) $Q^c$, the sum of all players' rewards, for the EBS, (2) $Q^r_j$, the maximum of each player's rewards given that all other players are minimizing that player's rewards, for punishment, and (3) $Q^d_j$, the maximum reward each player could achieve by deviating while all others adhere to the EBS. With this third value, an FCL player can compute the minimum number of episodes of punishment necessary to make the EBS the deviating player's optimal response.

The key insight is that if all users of FCL explore at the same times, they will cover the state and action spaces such that all three classes of action-value estimates $\hat{Q}$ eventually converge to the truth. This coordination of exploration both prevents misinterpretation of exploratory actions as deviations, and ensures that $\hat{Q}$ will accurately estimate cumulative returns under the EBS policy. The EBS and punishment policies maximize and minimize $Q^c$ and $Q^r_j$, respectively, where $j$ is the deviating player.

\section{Conclusion}

Our examination of the self-play compatibility problem in MAL offers two key contributions. First, we have provided clarity on the tradeoffs that arise based on the algorithm designers' choice of evaluation criteria. Unlike single-agent RL, it is not immediately clear what ``optimality'' is in self-play, as this depends on the standards of the other users one hopes will adopt one's algorithm. We see from the literature that although Pareto efficiency, bargaining solutions, and learning equilibrium are closer to ideal for self-play than attaining any Nash equilibrium, the latter is easier to balance with other-play guarantees. Second, this survey can serve as an organized reference for researchers aiming to develop incentive-stable algorithms. 

Several open problems remain despite a trend in self-play MAL towards the more ambitious solution concepts. In games where agents do not know each other's reward functions upfront, how can a single algorithm guarantee Pareto efficient bargains with its copies, while also adapting to other classes of agents? Both the S-Algorithm \cite{satisficing} and UCRG \cite{TD20}, suited to the setting of unknown rewards, cannot guarantee the latter property. Experimental results for M-Qubed \cite{CG10} suggest that this balance is feasible, but without a rigorous theoretical explanation or guarantee, the robustness of M-Qubed's approach to more complex settings and shorter time horizons is in doubt. 

Yet another useful extension to many of the algorithms introduced is how to assess and respond to a so-called targeted adversary. That is, do there exist specific cases for an algorithm for which the algorithm performs well below optimal levels? For instance in FCL \cite{fcl}, if two copies of the algorithm are pushing toward ``opposite'' cooperative solutions, they may believe that the other is continuously defecting. As a result, the algorithms continuously punish one another to bring each other in line with their desired strategy, resulting in a punishment loop that could be badly sub-optimal. Therefore, it may be useful to specify additional criteria to identify and avoid these deleterious situations. 

The EBS emerges from an egalitarian perspective on fairness (e.g. the work of philosopher John Rawls). Still, there are many other notions of fairness that merit exploration in Markov games. For instance, one could begin from a utilitarian perspective where the former's ``principle of equal gains'' is replaced by the ``principle of greatest good.'' Other notions similar to the NBS include the Kalai-Smorodinsky bargaining solution, which gives different guarantees. Lastly, it may also be interesting to design algorithms which can identify malicious agents in games where cooperation is mutually beneficial and accordingly adjust their perspective of fairness.

\newpage

\bibliographystyle{abbrv} 
\bibliography{self_play} 

\newpage

\section{Appendix}

\begin{table}[h]
    \centering
    \begin{tabular}{ |c|c|c| } 
 \hline
  & P2 Stag & P2 Rabbit \\ \hline
 P1 Stag & 5,5 & 0,1 \\ \hline
 P1 Rabbit & 1,0 & 1,1 \\ 
 \hline
\end{tabular}
    \caption{Stag Hunt}
    \label{tab:sh}
\end{table}

\vspace{30mm}

\begin{table}[h]
    \centering
\begin{tabular}{ |c|c|c| } 
 \hline
  & P2 Cooperate & P2 Defect \\ \hline
 P1 Cooperate & 3, 3 & 0, 4 \\ \hline
 P1 Defect & 4, 0 & 1, 1 \\ 
 \hline
\end{tabular}
    \caption{Prisoner's Dilemma}
    \label{tab:pd}
\end{table}

\begin{figure}[!b]\label{EBS}
  \centering
    \includegraphics[width=0.35\textwidth]{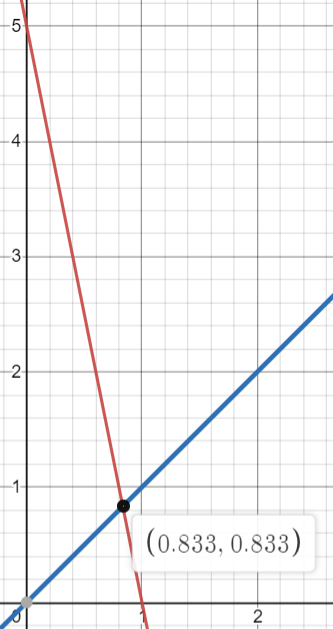}
    \caption{EBS solution in example}
\end{figure}

\end{document}